\documentstyle[twoside,fleqn,espcrc2]{article}

\catcode`\@=11 
\def\coeff#1#2{{\textstyle{#1\over #2}}}

\def\vev#1{\left\langle #1\right\rangle}
\def\lsim{\mathrel{\mathpalette\@versim<}}
\def\gsim{\mathrel{\mathpalette\@versim>}}
\def\@versim#1#2{\vcenter{\offinterlineskip
    \ialign{$\m@th#1\hfil##\hfil$\crcr#2\crcr\sim\crcr } }}

\def\JL{J. L. Lopez}
\def\DVN{D. V. Nanopoulos}
\def\r#1{$\bf#1$}
\def\rb#1{$\bf\overline{#1}$}
\def\y{\,{\rm y}}

\def\MeV{\,{\rm MeV}}
\def\GeV{\,{\rm GeV}}

\def\F{\widetilde F}
\def\Fb{\widetilde{\bar F}}
\def\NPB#1#2#3{Nucl. Phys. B {\bf#1} (19#2) #3}
\def\PLB#1#2#3{Phys. Lett. B {\bf#1} (19#2) #3}
\def\PRD#1#2#3{Phys. Rev. D {\bf#1} (19#2) #3}
\def\PRL#1#2#3{Phys. Rev. Lett. {\bf#1} (19#2) #3}

\newcommand{\AmS}{{\protect\the\textfont2
  A\kern-.1667em\lower.5ex\hbox{M}\kern-.125emS}}

\title{String Unification and Leptophobic $Z'$ in Flipped SU(5)}

\author{Jorge L. Lopez\address{Bonner Nuclear Lab, MS 315, Rice University,
6100 Main Street, Houston, TX 77005}
\thanks{To appear in Proceedings of Fourth International Conference on
Supersymmetry (SUSY96). Work supported in part by DOE grant DE-FG05-93-ER-40717.}}
       
\begin{document}
\begin{abstract}
We summarize recent developments in the prediction for $\alpha_s(M_Z)$,
self-consistent string unification and the dynamical determination of mass scales, and leptophobic $Z'$ gauge bosons in the context of stringy flipped SU(5).
\end{abstract}

\maketitle

\section{An oldie but goodie}
Flipped SU(5) enthusiasts keep discovering hidden treasures, even after 
10 years from its birth \cite{Erice95}. As is well known, the model attains its highest relevance in strings: efforts by several groups using different approaches have not (yet?) yielded appealing ``string GUTs" [SO(10)]. Among level-one Kac-Moody models, only flipped SU(5) unifies SU(3) and SU(2), providing an explanation for the ``LEP scale" [$10^{16}\GeV$]. The discrepancy between ``observed" and
predicted unification scales -- $M_{\rm LEP}\sim10^{16}\GeV$ versus $M_{\rm string}\sim10^{18}\GeV$ -- seems to have only way out: extra intermediate-scale states \cite{DF}. This solution was
realized early on in stringy flipped SU(5) \cite{search}. Here we
summarize how this scenario may be achieved in practice \cite{TwoStep}, including the prediction for $\alpha_s(M_Z)$ \cite{Lowering}, and also discuss the latest ``flipped" goodie: a leptophobic $Z'$ \cite{Zprime}.

\section{Some basics first}
\noindent \underline{Matter fields}:\\
$F_{(10)}=\{Q,d^c,\nu^c\}$; $\bar f_{(\bar 5)}=\{L,u^c\}$; $l^c_{(1)}=e^c$ $(\times 3)$\\
$F_{(10)}=\{Q,d^c,\nu^c\}$; 
$F_{(\overline{10})}=\{\bar Q,\bar d^c,\bar\nu^c\}$\\
\underline{Higgs fields}:\\
$H_{(10)}=\{Q_H,d^c_H,\nu^c_H\}$; $\bar H_{(\overline{10})}=\{Q_{\bar H},d^c_{\bar H},\nu^c_{\bar H}\}$ 
 $h_{(5)}=\{H_2,H_3\}$; $\bar h_{(\bar5)}=\{\bar H_2,\bar H_3\}$\\
\underline{GUT superpotential}:
\[
W_G=H\cdot H\cdot h+\bar H\cdot\bar H\cdot\bar h+F\cdot\bar H\cdot\phi
+\mu h\bar h
\]  
The vevs $\vev{\nu^c_H}=\vev{\nu^c_{\bar H}}=M_U$ break $SU(5)\times U(1)$
down to $SU(3)\times SU(2)\times U(1)$. \\
\begin{flushright}
\vspace*{-5.75cm}
DOE/ER/40717-30\\
\tt hep-ph/9607231
\vspace{4.cm}
\end{flushright}
\underline{Doublet-triplet splitting}: 
\[
h=\left(
\begin{array}{c}
H_2\\
H_3
\end{array}
\right)
\begin{array}{l}
{\rm electroweak\ symmetry\ breaking}\\
{\rm mediates\ proton\ decay}
\end{array}
\]
\begin{eqnarray}
&H\cdot H\cdot h\to d^c_H\vev{\nu^c_H}H_3\nonumber\\
&\bar H\cdot\bar H\cdot\bar h\to \bar d^c_H\vev{\bar\nu^c_H}\bar H_3\nonumber
\end{eqnarray}
The triplets get heavy, while the doublets remain light (``missing
partner mechanism").\\
\underline{Yukawa superpotential}:
\[
\lambda_d F\cdot F\cdot h+\lambda_u F\cdot\bar f\cdot \bar h+\lambda_e
\bar f\cdot l^c\cdot h
\]
\underline{Neutrino masses}: The GUT couplings $F\cdot \bar f\cdot h\to m_u\nu\nu^c$, $F\cdot \bar H\cdot\phi\to \vev{\nu^c_{\bar H}}\nu^c\,\phi$ entail
\[
M_\nu=
\begin{array}{c}
\nu\\ \nu^c\\ \phi
\end{array}
\stackrel{\begin{array}{ccc} \nu\quad&\nu^c&\quad\phi\end{array}}
{\left(
\begin{array}{ccc}
0&m_u&0\\ m_u&0&M_U\\ 0&M_U&M
\end{array}
\right)}
\]
See-saw mechanism: $m_{\nu_{e,\mu,\tau}}\sim m^2_{u,c,t}/(M_U^2/M)$\\
Good for MSW mechanism, $\nu_\tau$ hot dark matter, and ($\nu^c$) baryogenesis.\\
\underline{Dimension-six proton decay}: mediated by $X,Y$ GUT gauge bosons, the
mode $p\to e^+\pi^0$ may be observable at SuperKamiokande.\\
\underline{Dimension-five proton decay}:
\includegraphics{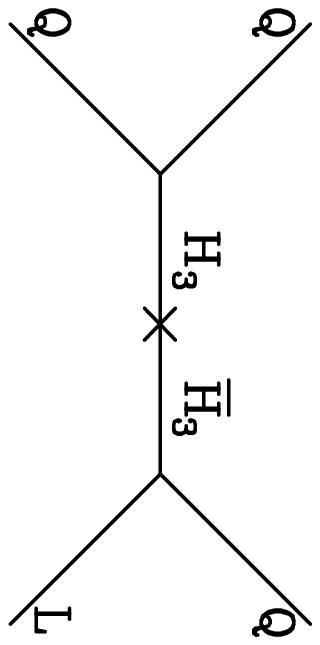}
very suppressed since no $H_3,\bar H_3$ mixing exists, even though $H_3,\bar H_3$ are heavy via doublet-triplet splitting, 
\[
\lambda_d F\cdot F\cdot h\supset QQH_3\qquad
\lambda_ u F\cdot\bar f\cdot\bar h\supset QL\bar H_3
\]
\newpage

\section{Prediction for $\alpha_s(M_Z)$}
\begin{figure}[t]
\vspace{5cm}
\includegraphics{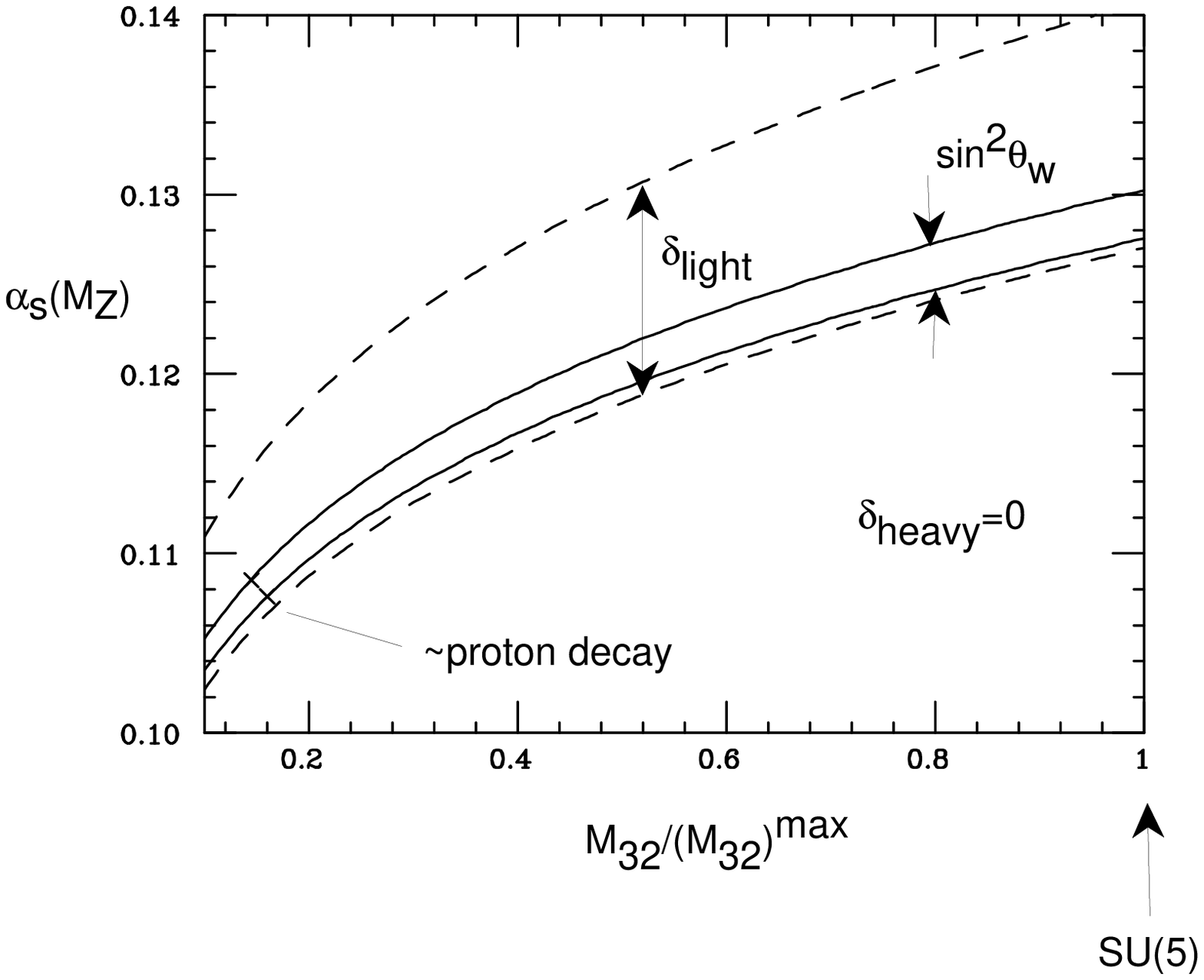}
\vspace{0cm}
\caption{Prediction for $\alpha_s(M_Z)$ versus the SU(2)/SU(3) unification scale $M_{32}$.}
\label{fig:sin2r}
\vspace{5cm}
\includegraphics{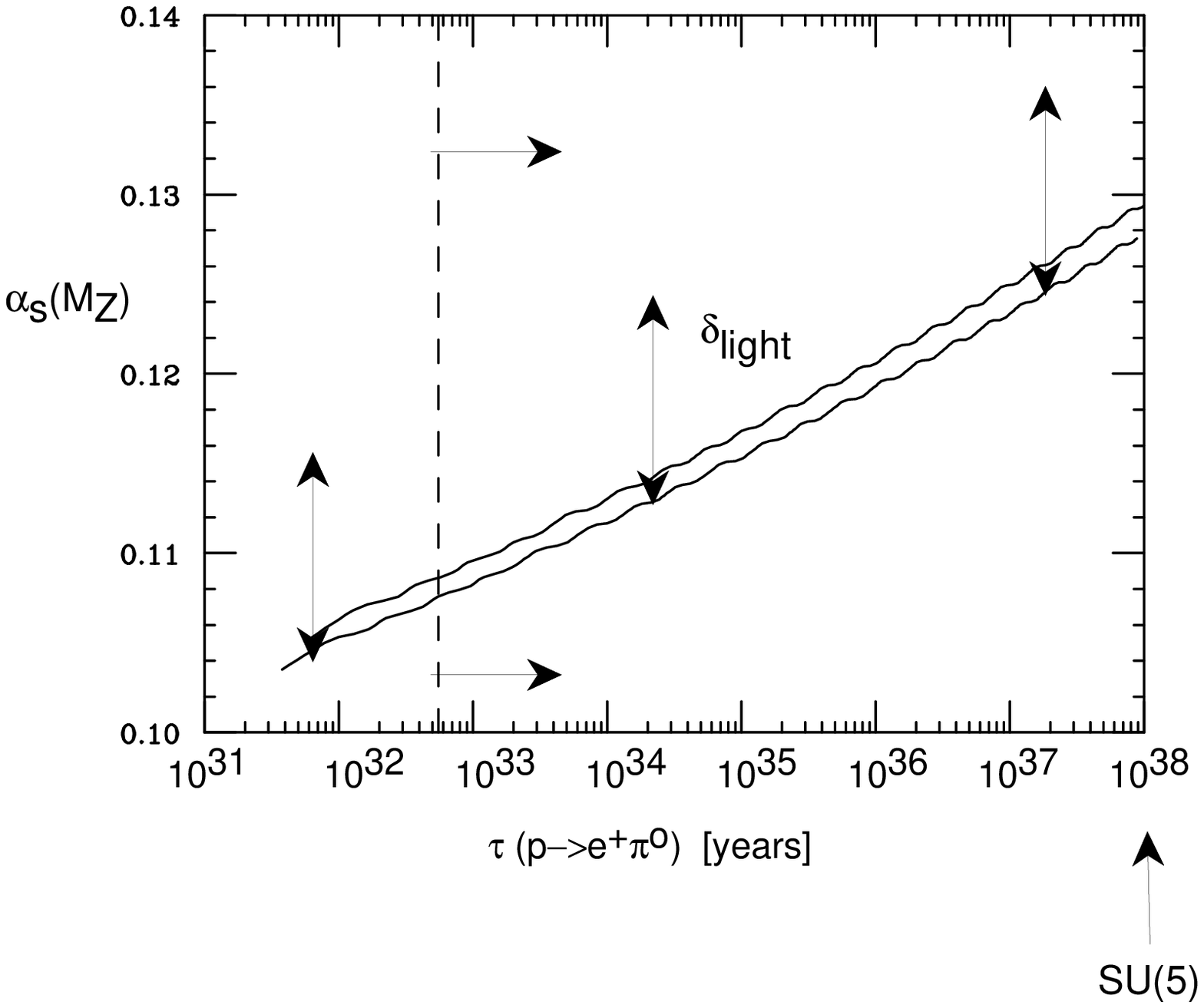}
\vspace{0.2cm}
\caption{Prediction for $\alpha_s(M_Z)$ versus the proton lifetime. Present lower bound indicated.}
\label{fig:taup}
\end{figure}

\noindent Starting from the low-energy Standard Model gauge couplings, and
evolving them from low to high energies, first $\alpha_2$ and $\alpha_3$ unify at $M_{32}$ \cite{Lowering}
\begin{eqnarray}
{1\over\alpha_2}-{1\over\alpha_5}&=&{b_2\over2\pi}\,
\ln{M_{32}\over M_Z}\nonumber\\
{1\over\alpha_3}-{1\over\alpha_5}&=&{b_3\over2\pi}\,
\ln{M_{32}\over M_Z}\nonumber
\end{eqnarray}
The hypercharge does not unify at $M_{32}$:
\[{1\over\alpha_Y}-{1\over\alpha_1'}={b_Y\over2\pi}\,\ln{M_{32}\over M_Z}\]
Above $M_{32}$ the gauge group is SU(5)$\times$U(1).
Stringy unification occurs at $M_{51}\ge M_{32}$ -- there is an $M_{32}^{\rm max}$
\[{1\over\alpha_Y}-{1\over\alpha_5}={b_Y\over2\pi}\,
\ln{M^{\rm max}_{32}\over M_Z}\]
Solving for $\alpha_3$, to lowest order:
\[\alpha_s(M_Z)={\coeff{7}{3}\,\alpha\over 5\sin^2\theta_W-1
+{11\over2\pi}\,\alpha\ln(M^{\rm max}_{32}/M_{32})}\]
compare with SU(5) where $M_{32}=M_{32}^{\rm max}$ \cite{Lowering}
\[ \alpha_s(M_Z)^{\rm Flipped\ SU(5)}<\alpha_s(M_Z)^{\rm SU(5)}\]
What happens at next-to-leading order?
\[\sin^2\theta_W\to \sin^2\theta_W-\delta_{\rm 2loop}
-\delta_{\rm light}-\delta_{\rm heavy}\]
Decreasing $\sin^2\theta_W$ increases $\alpha_s(M_Z)$ [avoid!]:
$\delta_{\rm 2loop}\approx0.0030$;
$\delta_{\rm light}\gsim0$ (light SUSY thresholds);
$\delta_{\rm heavy}$ (GUT thresholds)
\[\delta_{\rm heavy}={\alpha\over20\pi}
\left[ -6\ln{M_{32}\over M_{H_3}}-6\ln{M_{32}\over M_{\bar H_3}}
+4\ln{M_{32}\over M_V}\right]\]
Since there is no problem with proton decay, $\delta_{\rm heavy}$ can be
negative. We obtain $\alpha_s(M_Z)$ as low as 0.108 (see Fig.~\ref{fig:sin2r}).
However, decreasing $M_{32}$ decreases the proton lifetime
\[\tau(p\to e^+\pi^0)\approx1.5\times10^{33}
\left({M_{32}\over10^{15}\GeV}\right)^4
\left({0.042\over\alpha_5}\right)^2 {\rm y}\]

The present lower bound $\tau(p\to e^+\pi^0)^{\rm exp}>5.5\times10^{32}\y$
implies $\alpha_s(M_Z)>0.108$ (see Fig.~\ref{fig:taup}). If $\alpha_s(M_Z)<0.114$ then $p\to e^+\pi^0$
may be observable at SuperKamiokande (which should have a sensitivity of
$\sim10^{34}\y$). This is in contrast with minimal
SU(5), where the preferred mode is $p\to\bar\nu K^+$.

\section{Stringy Flipped SU(5)}
\noindent String construction in fermionic formulation \cite{search}\\
\underline{Gauge group}: 
$G=G_{\rm observable}\times G_{\rm hidden}\times G_{\rm U(1)}$ \\ 
$G_{\rm observable}$=SU(5)$\times$U(1);
$G_{\rm hidden}$=SO(10)$\times$SU(4);
$G_{\rm U(1)}$=U(1)$^5$\\
\underline{Particle spectrum}\\
Observable Sector:
\[
\begin{array}{lll}
F^{\{0,1,2,3,4\}}[10]
&\bar f^{\{2,3,5\}}[\bar 5]&{\ell^c}^{\{2,3,5\}}[1]\\
\bar F^{\{4,5\}}[\overline{10}]\\
h^{\{1,2,3,45\}}[5]&\bar h^{\{1,2,3,45\}}[\bar5]
\end{array}
\]
Singlets: 20 charged under U(1)'s, 4 neutral 
\newpage
\noindent Hidden Sector:
\[
\begin{array}{ll}  
\rm T^{\{1,2,3\}}& \rm [10]\ of\ SO(10)\\
\rm D^{\{1,2,3,4,5,6,7\}}&\rm [6]\ of\ SU(4)\\
\rm \F^{\{1,2,3,4,5,6\}}& \rm [4]\ of\ SU(4)\\
\rm \Fb^{\{1,2,3,4,5,6\}}& \rm [\bar4]\ of\ SU(4)
\end{array}
\]
The $\F_i,\Fb_j$ fields carry $\pm1/2$ electric charges
and exist only \underline{confined} in hadron-like {\em cryptons}.

The cubic and non-renormalizable terms in the superpotential have been
calculated \cite{search}, and more recently also the K\"ahler potential 
\cite{LN}. The properties of the K\"ahler potential illuminate the vacuum
energy (which vanishes at tree level and possibly also at one loop) and
determine the pattern of soft-supersymmetry-breaking masses, which has distinct experimental consequences \cite{LNZ}.

\section{String unification}
\noindent Assume that 
\[\rm SU(5)\times U(1)\to SU(3)\times SU(2)\times U(1)\]
breaks as in Standard Flipped SU(5) case.
Cancellation of $\rm U_A(1)$ is consistent with
\[ M_{\rm LEP}\sim\vev{\nu^c_H}\sim10^{15-16}\GeV\]
Correct $\sin^2\theta_W$ and $\alpha_3$ obtained because of extra
\r{10},\rb{10} present in string massless spectrum.
String unification occurs at $M_{\rm string}\sim10^{18}\GeV$.
This requires $M_{10}\sim10^{8-9}\GeV$, which can be generated via VEVs of hidden matter fields. 

\begin{figure}[b]
\includegraphics{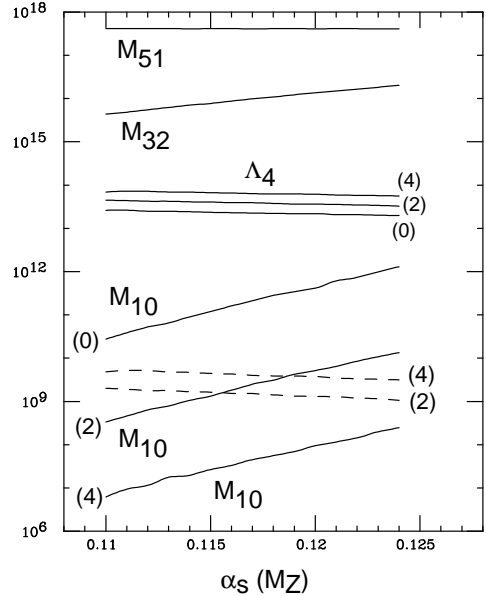}
\caption{The calculated values of $M_{51}=M_{\rm string}$, $M_{32}=M_{\rm LEP}$, $\Lambda_4$, and $M_{10}$, as a function of $\alpha_s(M_Z)$ for $N_4=0,2,4$ (indicated in parenthesis). Dashed lines display estimates of the dynamical prediction for $M_{10}$.}
\label{fig:scales}
\end{figure}

\begin{figure}[t]
\vspace{5cm}
\includegraphics{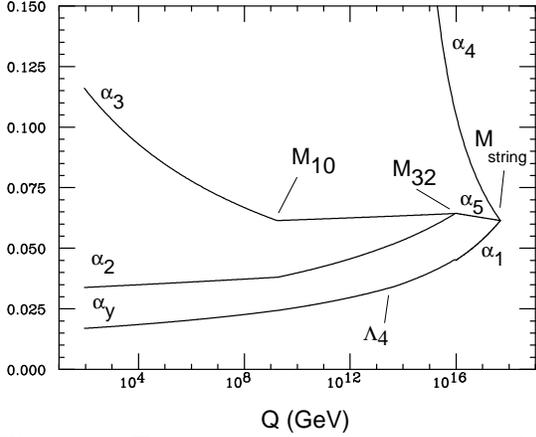}
\caption{The running of the gauge couplings for
$\alpha_s(M_Z)=0.116$ and $N_4=2$. One obtains $M_{10}=1.8\times10^9$~GeV,
$M_{32}=8.7\times10^{15}$~GeV, $M_{51}=4.4\times10^{17}$~GeV, $\Lambda_4=3.9\times10^{13}$~GeV, and $g=0.88$. This value of $M_{10}$ agrees
rather well with the dynamical prediction $M_{10}\sim10^9$~GeV.}
\label{fig:runnings}
\end{figure}

\noindent \underline{Dynamical Determination of Scales} \cite{TwoStep}\\
$\bf4,\bar{\bf4}$ affect running of U(1) down to
$\Lambda_4=M_{\rm string}\ e^{8\pi^2/g^2\beta_4}$, where
$\beta_4=-12+{1\over2}N_4+N_6$; $\Lambda_4$ depends on string spectrum of
$\bf4,\bar{\bf4},\bf6$, and on actual decoupling of particles between
$M_{\rm string}$ and $\Lambda_4$ [tricky]. Extra
$\bf10,\overline{\bf10}$ affect running of SU(5)$\times$U(1) down to $M_{10}$.
Naively, if $M_{4,\bar4}\sim\Lambda_4$, a non-renormalizable term
$(10)(\overline{10})(4)(\bar4){1\over M}$ implies 
$M_{10}\sim {\vev{4\bar 4}\over M}\sim {\Lambda^2_4\over M}$.
But in strings $M_{4,\bar4}\sim\Lambda_4\ll M$ is very unlikely;
$M_{4,\bar4}=0$ is more natural. In the actual string model we have a quintic term [and $M_{4,\bar4}=0$]: $(10)(\overline{10})(4)(\bar4)\phi{1\over M}$,
where the cancellation of $\rm U_A(1)$ implies $\vev{\phi}/M\sim1/10$.
With massless flavors ($M_{4,\bar4}=0$) one expects $\vev{4\bar4}\sim \infty$.
Aharony, et. al. studied $\rm SU(N_c)$ with $N_f$ ``massless" flavors with supersymmetry-breaking scalar masses  \cite{Peskin}.
Supersymmetry-breaking masses ($\tilde m$) yield finite condensates
\[ \vev{H\bar H}\sim \left[{N_c\over N_c-N_f}\ {1\over \tilde m}\right]
^{(N_c-N_f)/2(2N_c-N_f)}\]
In our case ($N_c=4$, $N_f=2$) we obtain
\[\vev{4\bar 4}\sim \Lambda^2_4\left({\tilde m\over \Lambda_4}\right)^{-1/3}
\gg\Lambda^2_4\ ,\]
and we can \underline{calculate} $M_{10}$ from first principles
\[ M_{10}\sim \left({\Lambda_4\over M}\right)^2
\left({\tilde m\over \Lambda_4}\right)^{-1/3}{\vev{\phi}\over M}\ M\sim
10^{8\to10}\GeV\]
This result allows self-consistent string unification. The 
results for the various scales as a function of $\alpha_s(M_Z)$ are shown
in Fig.~\ref{fig:scales}. The full evolution of the gauge couplings from the weak scale to the string scale is shown in Fig.~\ref{fig:runnings} for the preferred choices of $\alpha_s(M_Z)=0.116$ and $N_4=2$.

\clearpage

\section{Leptophobic $Z'$}
\noindent\underline{Motivation}:
Original ``smoking gun" of string; $R_b,R_c$ `crisis' has revived interest in $Z'$ models, although this time the $Z'$ must not couple to leptons. Leptophobia is {\em natural} in flipped SU(5) \cite{Zprime}
\[ 10=\{Q,d^c,\nu^c\};\quad \bar 5=\{L,u^c\};\quad 1=e^c\]
If the leptons are uncharged, most quarks may be charged under $\rm U'$. Compare with regular SU(5) $10=\{Q,u^c,e^c\};\quad \bar 5=\{L,d^c\}$,
where uncharged leptons imply uncharged quarks. Dynamic leptophobia (via RGE U(1) mixing) is also possible, as in the $\eta$-model in $E_6$ \cite{BKM}.

Any $Z$-$Z'$ mixing shifts the usual $Z$ couplings ($C^0_{V,A}$):
$C_V=C_V^0+\theta(g_{Z'}/g_Z)\,C'_V$, $C_A=C_A^0+\theta(g_{Z'}/g_Z)\,C'_A$,
where $\theta$ is the $Z$-$Z'$ mixing angle (small); $g_{Z},g_{Z'}$ are the $Z,Z'$ gauge couplings; and $C'_{V,A}$ the fermion couplings to the $Z'$.
In flipped SU(5)  we have [$C'_V=Q_L+Q_R$, $C'_A=-Q_L+Q_R$]
\[
\begin{tabular}{c|crcccc}
&$C^0_V$&$C^0_A$&$Q_L$&$Q_R$&$C'_V$&$C'_A$\\ \hline
$u$&${1\over2}-{4\over3}x_w$&${1\over2}$&$c$&0&$c$&$-c$\\
$d$&$-{1\over2}+{2\over3}x_w$&$-{1\over2}$&$c$&$c$&$2c$&0\\
\end{tabular}
\]
We can determine the first-order shifts in $\Gamma_{c\bar c}$, $\Gamma_{b\bar b}$, and $\Gamma_{\rm had}$, allowing for non-universal $c_{1,2,3}$ charges picked from
\[
\begin{tabular}{lrclrclr}
$F_0$&$-\coeff{1}{2}$&\qquad&$\bar F_4$&${1\over2}$&\qquad&$\bar f_{2,3,5}$&0\\
$F_1$&$-\coeff{1}{2}$&&$\bar F_5$&0&&$\ell^c_{2,3,5}$&0\\
$F_2$&0\\
$F_3$&1\\
$F_4$&$-\coeff{1}{2}$
\end{tabular}
\]
This $\rm U'$ charge space satisfies specific requirements:
The leptons (in $\bar f_{2,3,5},\ell^c_{2,3,5}$) are uncharged; one uncharged $({\bf10},\overline{\bf10})$ pair ($F_2,\bar F_5$) so that $\rm U'$ remains unbroken upon SU(5)$\times$U(1) breaking; $\rm Tr\,U'=0$ enforced; extra $({\bf10},\overline{\bf10})$ to allow string unification. The actual string model underlies these choices. 

\begin{figure}[b]
\includegraphics{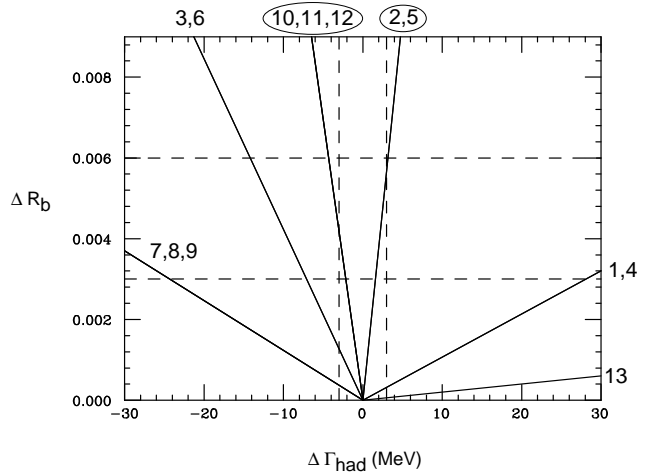}
\vspace{0.8cm}
\caption{Correlated shifts in $R_b$ and $\Gamma_{\rm had}$ for the various $\rm U'$ charge assignment combinations. Dashed lines delimit the experimental limits on $\Delta\Gamma_{\rm had}$ and $\Delta R_b$. Circled charge assignments (2,5,10,11,12) agree with experiment.}
\label{fig:RbGhad}
\end{figure}

There are 13 possible charge assignments that can be made. Phenomenology demands $\Delta\Gamma_{\rm had}\lsim 3\MeV$, as the SM prediction and LEP agree well. Since $R^{\rm SM}_b=0.2157$ and $R^{\rm exp}_b=0.2202\pm0.0016$
($R_c$ fixed to SM value), we demand $\Delta R_b=0.0030-0.0060$. Fig.~\ref{fig:RbGhad} shows $\Delta R_b$ versus $\Delta\Gamma_{\rm had}$.
An analogous plot for $\Delta R_c$ versus $\Delta R_b$, demanding
$\Delta R_c$, $\Delta R_b$ shifts in opposite directions can be found in Ref.~\cite{Zprime}. We should keep in mind that experimentally there appears
to be a trend of $R_c$ converging to the Standard Model prediction and
$R_b$ approaching it significantly.

\clearpage

\subsection{String scenario}
\noindent Consider $G_{\rm U(1)}=\rm U_1\times U_2\times U_3\times U_4\times U_5$, with $\rm Tr\,U_4=0$, $\rm Tr\,U_{1,2,3,5}\not=0$. The anomalous
combination is $\rm U_A=U_1-3U_2+U_3+2U_5$, with three orthogonal traceless combinations: $\rm U'_1=U_3+2U_5$; $\rm U'_2=U_1-3U_2$;
$\rm U'_3=3U_1+U_2+4U_3-2U_5$.
The lepton charges under $\rm U'_1,U'_2,U'_3$ are
$\bar f_{2,5},\ell^c_{2,5}:(0,{3\over2},-{1\over2})$;
$\bar f_3,\ell^c_3:({3\over2},0,1)$.\\
There is a {\em unique} $\rm U'$ that is leptophobic
\[ \rm U'\propto 2U'_1-U'_2-3U'_3
\propto U_1+U_3-U_5\]
and by construction $\rm Tr\, U'=0$.
Higgs fields charged under $\rm U'$ exist ($Z$-$Z'$ mixing).
The D- and F-flatness conditions may be satisfied, leaving $\rm U'$
unbroken, but breaking the hidden group.\\
\underline{Model building}: $F_4$ should contain 3rd generation (top Yukawa);
$F_2,\bar F_5$ neutral under $\rm U'$: symmetry breaking;
$\bar F_4$: string unification; $R_b,R_c$ inputs: four charge assignments allowed 
\[\begin{tabular}{c|rrr}
&$c_1$&$c_2$&$c_3$\\ \hline
(2)&0&$-{1\over2}$&1\\
(5)&$-{1\over2}$&0&1\\
(11)&$-{1\over2}$&$1$&$-{1\over2}$\\
(12)&$-{1\over2}$&$-{1\over2}$&$1$
\end{tabular}
\]
Unlike any considered before. Unnatural? Obtained from string!
Top-quark Yukawa coupling, and $R_b,R_c$ select scenario (11) uniquely
\[\Delta R_b\approx 0.0042\,
\left({\Delta\Gamma_{\rm had}\over -3\,{\rm MeV}}\right)\,,\quad
\Delta R_c\approx -0.76\,\Delta R_b\ .\]
\underline{Dynamics}: Running of $\rm U'$ from $M_Z$ up looks good:
$b'={16\over3}$ (c.f. $b_Y={33\over5}$). Sufficiently small $Z$-$Z'$ mixing
appears to require radiative $\rm U'$ symmetry breaking via singlet $\vev{\phi}$.

\subsection{Experimental prospects}
\noindent $Z'$ width and branching ratios for preferred case:
\[
\begin{tabular}{c|rrc}
(11)&$C'_V$&$C'_A$&$B(Z'\to q\bar q)$\\ \hline
$u$&$-{1\over2}$&$-{1\over2}$&$1\over18$\\
$d$&$-1$&0&$1\over9$\\
$c$&1&1&$2\over9$\\
$s$&2&0&$4\over9$\\
$t$&$-{1\over2}$&$-{1\over2}$&$1\over18$\\
$b$&$-1$&0&$1\over9$
\end{tabular}
\]
\[{\Gamma_{Z'}\over M_{Z'}}\approx 0.033 \left({g_{Z'}\over g_Z}\right)^2
\ {\rm[narrow]}\]
Experimental limits: 
\begin{eqnarray}
{\widehat\sigma(u\bar u\to Z')\over \widehat\sigma(u\bar u\to Z')_{\rm SM}}
&\approx& 0.58
 \left({g_{Z'}\over g_Z}\right)^2\nonumber\\
{\widehat\sigma(d\bar d\to Z')\over \widehat\sigma(d\bar d\to Z')_{\rm SM}}
&\approx& 0.90
\left({g_{Z'}\over g_Z}\right)^2\nonumber
\end{eqnarray}
Average up/down; multiply by {${B(Z'\to jj)\over B(Z'\to jj)_{\rm SM}}\approx1.4$},
\[\sigma(p\bar p\to Z'\to jj)\approx \left({g_{Z'}\over g_Z}\right)^2 \sigma(p\bar p\to Z'\to jj)_{\rm SM}\]
Only limit from UA2: $M_{Z'}>260\GeV$, but only if $g_{Z'}=g_{Z}$.

$Z'$ contributes to top-quark cross section (see Fig.~3 in Ref.~\cite{Zprime})
at a level that may be observable if $M_{Z'}\sim500\GeV$. Parity-violating spin asymmetries at RHIC may also show deviations from Standard Model expectations
because of the $t$-channel exchange of our parity-violating $Z'$.

In sum, flipped SU(5) continues to provide unsolicited solutions to unanticipated problems, as evidenced most recently by the self-consistent
string unification and the possible existence of a leptophobic $Z'$ gauge
boson.

\end{document}